\documentclass[12pt]{article}
\usepackage{amsmath,amssymb,amsfonts}
\usepackage[dvips]{graphicx}
\usepackage{epsfig}

\makeatletter \@addtoreset{equation}{section} \makeatother
\makeatletter \@addtoreset{figure}{section} \makeatother

\def\CA{{\cal A}}
\def\CF{{\cal F}}
\def\CI{{\cal I}}\def\CJ{{\cal J}}
\def\CK{{\cal K}}\def\CL{{\cal L}}\def\CM{{\cal M}}
\def\CN{{\cal N}}\def\CP{{\cal P}}

\def\CW{{\cal W}}
\def\CZ{{\cal Z}}

\def\g{\gamma}
\def\d{\delta}\def\e{\epsilon}
\def\z{\zeta}

\begin{document}
\begin{titlepage}
\vfill
\begin{flushright}
{\tt\normalsize KIAS-P12054}\\

\end{flushright}
\vfill
\begin{center}
{\large\bf  Constructive Wall-Crossing and Seiberg-Witten}\footnote{This
article is a conference proceeding contribution for
{\it Progress of Quantum Field Theory and String Theory}, Osaka, April 2012.}

\vskip 1cm

Piljin Yi
\vskip 3mm
{\it School of Physics, Korea Institute
for Advanced Study, Seoul 130-722, Korea}

\end{center}
\vfill

\begin{abstract}
\noindent
We outline a comprehensive  and first-principle solution to the wall-crossing
problem in $D=4$ $N=2$ Seiberg-Witten theories. We start with a brief review of the
multi-centered nature of the typical BPS states and recall how  the wall-crossing 
problem thus becomes really a bound state
formation/dissociation problem. Low energy dynamics for arbitrary collections
of dyons is derived, from Seiberg-Witten theory, with the proximity to the so-called marginal stability
wall playing the role of the small expansion parameter. We find that, surprisingly, the
$\mathbb{R}^{3n}$ low energy dynamics of $n+1$ BPS dyons cannot be consistently 
reduced to the  classical moduli space, $\CM$, yet the index can be phrased in terms of $\CM$. 
We also explain how an equivariant version of this index computes the protected 
spin character of the underlying field theory, where $SO(3)_\CJ$ isometry of
$\CM$ turns out to be the diagonal subgroup of $SU(2)_L$ spatial rotation and
$SU(2)_R$ R-symmetry. The so-called rational invariants, previously seen in the 
Kontsevich-Soibelman formalism of wall-crossing, are shown to emerge  naturally 
from the orbifolding projection due to Bose/Fermi statistics.

\end{abstract}

\vfill
\end{titlepage}

\parskip 0.1 cm
\tableofcontents\newpage
\renewcommand{\thefootnote}{\#\arabic{footnote}}
\setcounter{footnote}{0}

\parskip 0.2 cm

\section{Multi-Center Picture and Wall-Crossing}	

Wall-crossing refers to phenomena where certain BPS states \cite{Prasad:1975kr,Bogomolny:1975de}
disappear or appear abruptly as  parameters or vacuum moduli of a supersymmetric
theory are continuously deformed. The places where this  happens define
the so-called marginal stability walls (MSW) that separate one domain from another
in the parameter/vacuum moduli space. Within each such domain,
the BPS spectrum is protected. In four dimensional
supersymmetric theories, the phenomenon was first discovered in $SU(2)$
Seiberg-Witten theory \cite{Seiberg:1994rs,Seiberg:1994aj}: In the weak coupling limit
the BPS spectrum is infinite, with the massive vector meson and certain infinite
tower of dyons, while in a strongly coupled regime one finds only two
particle-like BPS states, namely a monopole and a unit-electrically charged dyon \cite{Ferrari:1996sv}.
Such phenomena turned out generic
for BPS states (say, of higher supersymmetric theories) that preserve four
or less supercharges.

Given the $SU(2)$ example with the MSW deep in the strongly coupled
regime, one might be mislead to think that  wall-crossing is inherently
strong coupling phenomena. However, nothing can be further from the truth.
An example where this can be seen most easily is the 1/4 BPS state of
$N=4$ super-Yang-Mills theories; these objects preserve the same number
of supersymmetry as $N=2$ BPS states and are also affected by wall-crossing.
For those who prefer a stringy picture, one could equivalently consider
a $(p,q)$ string web with end points at D3-branes \cite{Bergman:1997yw}.
With the latter, one finds
various three-way junctions where three different types of $(p,q)$ strings
meets in a supersymmetric fashion. The angles between strings at such junctions
are determined entirely by balance of string tensions, or more precisely by
BPS conditions, and so are independent of how long each string segments are.
The latter means that one would inevitably encounter a MSW as a D3 is brought
near a junction shortening a string segment. Passing the D3 through that
junction violates the BPS condition, and a wall-crossing occurs. Because the
underlying theory has $N=4$ supersymmetry, one can choose to perform this
analysis with arbitrarily small coupling, which clearly shows that the
phenomenon must have a simple classical or semiclassical interpretation.

Understanding the wall-crossing of such states from the spacetime viewpoint, i.e., from
$N=4$ field theory viewpoint, came shortly afterward \cite{Lee:1998nv}.
It was shown that a typical 1/4 BPS soliton is made up of two or more well-separated
non-Abelian charge cores. Distances between these charge cores are determined
entirely by classical balance of forces, such that, as
vacuum moduli (equivalently, positions of D3) cross a MSW, at least one of
these distances diverges. At quantum level, this translate to divergent
size of the bound state wavefunction. With the bound state size no longer
square normalizable, one can no longer interpret the BPS state in question
as a single particle state. This is how BPS states disappear from the spectrum
across an MSW. All 1/4 BPS states are loose bound states of some simpler
subset of BPS particles, and at MSW's become classically destabilized \cite{Lee:1998nv,Bak:1999da}.

This multi-center picture and wall-crossing for 1/4 BPS states in $N=4$
Yang-Mills theories
was later extended to $N=2$ theories \cite{Gauntlett:1999vc,Gauntlett:2000ks},
where the same phenomenon was found:  $N=2$ BPS states are typically
bound states of a simpler class of BPS particles, whose wavefunctions
become nonnormalizable at MSW's. An index theorem, counting the discontinuity,
was also found and computed \cite{Stern:2000ie} for both $N=4$ 1/4 BPS
states and $N=2$ BPS states. While this development could not rigorously
address strongly coupled regimes, it was obvious that the intuitive
wall-crossing based on the multi-center picture should hold
universally and that the multi-centered nature of the BPS state
should become semiclassically manifest when a relevant MSW is approached.

Rediscovery of the same multi-centered nature in the context of BPS
black holes \cite{Denef:2000nb} allowed more systematic study
of the wall-crossing phenomena. Two main results emerged from this
and dominated the topic for a decade since.
The first offers a simple sets of constraints for relative
positions of charge cores. With charge  $\gamma_A$ at $\vec x_A$,
these positions are constrained as
\begin{eqnarray}\label{denef}
\sum_{B\neq A}\frac{\langle \gamma_B,\gamma_A\rangle}{|\,\vec x_B-\vec x_A|}
=2\,{\text{Im}\big[ \zeta_T^{-1} Z(\g_A) \big]}\ ,
\end{eqnarray}
where $Z(\g_A)$ is the central charge of $\gamma_A$
and $\zeta_T $ is the phase factor of the total central charge
$Z_T=\sum_A Z(\g_A)$. This formula by itself does not really
inform us how to decompose the total charge into $\sum_A\g_A$.
Nevertheless, once the latter is known, this tightly constrains
possible classical solutions.  Note how the quantities here are expressed entirely
in terms of charges and central charges, and other details of the
underlying theories drop out. The simplicity of the formula is in part
due to Abelian nature of black holes, but can be extended to
non-Abelian field theory solitons, as will be one of main point
of this talk, as long as we move near an MSW.
The second is an index  formula that count supersymmetric
bound states of two charges \cite{Denef:2002ru,Denef:2007vg}
\begin{equation}
\Omega^-(\g_1+\g_2)=(-1)^{|\langle \g_1,\g_2\rangle|-1}|\langle \g_1,\g_2\rangle|\,
\Omega^+(\g_1)\Omega^+(\g_2) \ ,
\end{equation}
where $\pm$ in $\Omega^\pm$ refers to the two sides of a MSW.
$\Omega$  is an index, called the 2nd helicity trace,
\begin{equation}
\Omega=-\frac12\,{\rm tr}\left((-1)^{2J_3}(2J_3)^2\right)\ , 
\end{equation}
whose value is $1$ for a half-hypermultiplet and $-2$
for a vector multiplet. More generally, for a BPS multiplet
consisting of spin $j$ angular momentum multiplet times a
half-hyper, the value is $(-1)^j(2j+1)$.
This so-called primitive wall-crossing formula has been observed
in many studies, generalized to the so-called semi-primitive cases
for $\g_1+k\g_2$ states \cite{Denef:2007vg}, and more
recently elevated to the Kontsevich-Soibelman
formalism \cite{KS,Gaiotto:2008cd,Gaiotto:2009hg,Chen:2010yr}.
Despite successes of these later and more comprehensive work,
much of the literature had remained mathematical, conjectural,
and, from physics viewpoint, rather opaque.

In this talk, we
explain a new a universal approach to BPS states in Seiberg-Witten
field theories \cite{Kim:2011sc}, whereby we can derive wall-crossing
formula in its full generality and entirely from the underlying
$D=4$ field theory. Combination of supersymmetry algebra and
the proximity to an MSW turns out to give us enough control to
constrain and solve low energy dynamics of generic Seiberg-Witten
dyons, and furthermore can be connected to more general framework
of quiver quantum mechanics \cite{Denef:2002ru}.
This new method, in addition to providing a rather
satisfactory and physical solution to the problem, also addresses
two outstanding issues in the wall-crossing literature.

The first is whether and under what circumstance the relatively
simple distance formula (\ref{denef})  can be extended for field
theory solitons with non-Abelian cores. Old field theory dynamics of BPS
dyons \cite{Bak:1999da,Gauntlett:1999vc,Gauntlett:2000ks,Weinberg:2006rq}
comes with similar distance formula for solitons; however, it
treats magnetic and electric charges differently, which
makes it ineffective for dyons with MSW located in the strongly
coupled regime. In our new approach, we start by addressing
this question; simply put, exactly the same distance formula
works, but only if when we approach an MSW; this proximity
to an MSW will also serve as a small parameter that controls the
low energy approximation \cite{Lee:2011ph,Ritz:2008jf} in the end.

The second concerns various issues surrounding the
quiver quantum mechanics formulation \cite{Denef:2002ru}. Although
very well motivated from wrapped D3 brane picture of BPS states,
the subsequent analysis lead to two questions.
One concerns what is the right index theorem to use in
the so-called Coulomb phase. The other is why the so-called
Higgs phase index agrees with the Coulomb phase only in some cases
and not in others. Both of these questions turns out to be
due to subtleties in the Coulomb phase of the quiver
quantum mechanics. For those who are more familiar with this
approach to wall-crossing, our work can be regarded as a physical
derivation and subsequent study of the Coulomb phase
viewpoint. We will end up addressing the first question head on,
in this talk, while referring the second to another, more recent study.

\section{Excursion: Kontsevich-Soibelman and Rational Invariants}

Before we derive physical wall-crossing formulae,
we wish to briefly digress to the algebraic formalism
by Kontsevich and Soibelman \cite{KS}. For this we start with Lie algebra generators
$e_\g$, one for each and every charge $\g$, such that their commutators are
\begin{equation}\label{Lie}
[e_{\g_1},e_{\g_2}]=(-1)^{\langle \g_1,\g_2\rangle}\langle \g_1,\g_2\rangle\, e_{\g_1+\g_2}\ ,
\end{equation}
with the Schwinger product $\langle \cdot,\cdot\rangle$.
This is then exponentiated to operators
\begin{equation}
K_{\g}=\exp\left(\sum_{n=1}^\infty \frac{e_{n\g}}{n^2}\right)\ .
\end{equation}
Given the indices $\Omega(\g_i)$, one then proceeds to build a product
of such operators as
\begin{equation}\label{string}
\cdots K_{\g_1}^{\Omega(\g_1)}\cdots K_{\g_2}^{\Omega(\g_2)}\cdots \ ,
\end{equation}
which encodes the details of the theory in question.

As these operators do not generally commute, the ordering is crucial
and dictated by the phase of central charges $Z(\g_i)$. While the
phase is a periodic variable and does not usually define an ordering,
there is a sense in which we may do so near an MSW. Since the phases
of wall-crossing states all line up there, we can order at least these
relative to their common phase at the MSW. Denoting again by the
superscript $\pm$ the two sides of an MSW where phases of
$Z(\g_1)$ and $Z(\g_2)$ line up, we have two  different
string of operators,
\begin{equation}
U^+=\cdots K_{\g_1}^{\Omega^+(\g_1)}\cdots K_{\g_2}^{\Omega^+(\g_2)}\cdots \ ,
\end{equation}
and
\begin{equation}
U^-=\cdots K_{\g_2}^{\Omega^-(\g_2)}\cdots K_{\g_1}^{\Omega^-(\g_1)}\cdots\ .
\end{equation}
There are also those states whose central charge phase does not line
up with $\g_{1,2}$ etc; for these with very different phases from any
associated with $n\g_1+k\g_2$, we  simply choose them to occupy the left
end or the right end of these operator products, or more to the point, to
occupy the same corner in $U^\pm$.

The KS conjecture states that the wall-crossing occurs to make sure that
these two strings of operators are actually identical in the end, i.e.,
$U^+=U^-$, even though the positions of individual $K$'s in $U^\pm$
are completely different. As positions of $K_{\g_{1,2}}$
are flipped, for instance, one must have different strings of $K_{n\g_1+k\g_2}$'s
between the two, in order for $U^+=U^-$ to hold; this comparison determines
$\Omega^-$, say, given $\Omega^+$'s. The equality of $U^\pm$ was later
interpreted as a continuity condition of the so-called Darboux coordinates
in compactified Seiberg-Witten theory \cite{Gaiotto:2008cd},
providing more physical motivation for the conjecture.
Obviously, one important check of this KS formalism is a comparison to
other wall-crossing formulae that are derived as a solution to
the original physics problem, such as ours. Recently  a complete agreement
between the two was found \cite{Sen:2011aa}
under the assumption that relevant $\g$'s belong
to a single plane through the origin in the charge lattice.
We will not repeat the proof, but instead point out one
very important ingredient.

Let us note that $\Omega(k\g)\neq 0$ with $k\neq1$ is a logical
possibility.
In fact, for BPS black holes, such non-primitive states would be routinely
expected.
Since $\g$ and $k\g$ share exactly the same central charge
phase, (\ref{string}) can be more precisely written as
\begin{equation}
\cdots \prod_{k_1} K_{k_1\g_1}^{\Omega(k_1\g_1)}\cdots
\prod_{k_2}K_{k_2\g_2}^{\Omega(k_2\g_2)}\cdots \ ,
\end{equation}
or equivalently as
\begin{equation}
\cdots V({\g_1})\cdots V({\g_2})\cdots \ .
\end{equation}
with $\g$'s now running only over the primitive charges; the operator
$V(\g)$ knows about the actual spectrum as \cite{Manschot:2010qz}
\begin{equation}
V^\pm(\g)\equiv \exp\left(\sum_{k=1}^\infty
\bar\Omega^\pm(k\g)\,e_{k\g}\right)\ , \qquad
\bar\Omega^\pm (\Gamma)\equiv \sum_{s\vert \Gamma}
\frac{\Omega^\pm (\Gamma/s)}{s^2}\ .
\end{equation}
The ``rational invariant" $\bar\Omega$ is defined
as a sum over positive integers $s$ such that $\Gamma/s$ is a
well-defined integral charge. The KS wall-crossing formula then says,
\begin{eqnarray}
U^+&=&\cdots V^+({\g_1})\cdots V^+(p_1\g_1+p_2\g_2)\cdots V^+({\g_2})\cdots \nonumber \\
&=&\cdots V^-({\g_2})\cdots V^-(p_1\g_1+p_2\g_2) \cdots V^-({\g_1})\cdots=U^-
\end{eqnarray}
with the product running over
primitive charges $\sum p_i\g_i$'s. From this, relationships between
the two sets of rational invariants $\bar\Omega^\pm(\sum n_i\g_i) $
emerge, and one can decode in favor of the physical quantities, $\Omega^\pm(\sum n_i\g_i)$'s.

This suggests that, if the KS formalism indeed provides
a universal answer to the wall-crossing problem, one should very
well expect that the rational invariants $\bar \Omega$ appear naturally in
any sensible derivation of the wall-crossing formula from physics
also, and in a manner that has nothing to do with details of
dynamics. As we will see shortly, the answer to this lies
in the quantum statistics.

\section{Low Energy Dynamics in the Strong Coupling Regime}

The first step is to consider a collection of $n+1$ charge $\gamma_A$'s
in Seiberg-Witten theory, and represent it as a semiclassical state.
The BPS equations of the Seiberg-Witten theory
is \cite{Chalmers:1996ya,Mikhailov:1998bx,Ritz:2000xa,Argyres:2001pv}
\begin{eqnarray}\label{BPS}
\vec F_I - i \z^{-1}_T \vec \nabla \phi_I = 0 \ ,\quad \vec F_{D}^I
  - i \z^{-1} \vec \nabla \phi^I_D =0 \ ,
\end{eqnarray}
where $F=B+iE$ with magnetic field $B$'s and electric field $E$'s, and
$\phi$'s are unbroken part of the complex adjoint scalars. They are
all labeled by the Cartan index $I=1,2,\dots,r$.
$F_D$'s are defined through the low energy $U(1)$ coupling matrix as
\begin{equation}
\vec F_D^I \equiv \tau^{IJ} \vec F_J \ , \qquad
\tau^{JI}=\frac{\partial \phi_D^J}{\partial\phi^I}\ .
\end{equation}
The pure phase factor $\zeta_T$ is determined by the
supersymmetry left unbroken by the total charge, namely
equals the phase factor of the total central charge,
$Z(\g_T)/|Z(\gamma_T)|$. Let us begin with  a
core-probe approximation, where we split
$\g_T=\g_h+\sum_{A'}\g_{A'}$
and treat the latter $n$ as a fixed background. Real parts of
$F$ and $F_D$ obeys a Bianchi/Gauss law

In this core-probe approximation, three quantities determine
entire low energy dynamics of the probe dyon, $\g_h$.
The first is the inertia function $f$ as in
\begin{equation}
\CL=\frac12\, f \left(\frac{d\vec x}{dt}\right)^2 +\cdots\ .
\end{equation}
The inertia $f$ is generally position-dependent because the
core dyons deform the scalar fields, which in turn deform the
effective local value of the central charge. Solving (\ref{BPS})
for a given spatial distribution of ``core charges" $\g_{A'}$
and constructing the relevant central charge function lead to \cite{Lee:2011ph}
\begin{equation}
\CZ_{\gamma_h}(\vec x) =\g^e_h\cdot\phi(\vec x)+\g^m_h\cdot\phi_D(\vec x)
 \ ,\qquad f(\vec x)=|\,{\cal Z}_{\g_h}(\vec x)|\ ,
\end{equation}
with the electric part $\g_h^e$ and the magnetic part $\g_h^m$ of
the probe charge vector $\g_{_h}$.

Note that, in construction of $\CZ_h$,
we do not include $\g_h$ as a charge source for $\phi$ and $\phi_D$;
for bound state construction we only need to understand how charge
centers affect other charge centers. Also $\phi$ and $\phi_D$
are approximate semiclassical configurations, as we
dropped the non-Abelian part by starting with Seiberg-Witten description.
However, this is good enough if individual charge centers are far apart
from each other and look effectively point-like. The latter is guaranteed,
on the other hand, by moving very near an MSW where relevant charge centers
separate far apart from one another.

The other two, more important quantities are the scalar potential
and the vector potential, respectively, ${\cal K}^2/2f$ and ${\cal W}$ in
\begin{equation}
\CL=\frac12\, f \left(\frac{d\vec x}{dt}\right)^2-\frac{\CK^2}{2f}-
\frac{d\vec x}{dt} \cdot\vec\CW+\cdots\ .
\end{equation}
A remarkable fact is that these two are again determined entirely by
the same central charge function  ${\cal Z}_{\g_h}$ as \cite{Lee:2011ph}
\begin{equation}
d{\cal W}= *d{\cal K}\ , \quad
{\cal K}={\rm Im}[\zeta^{-1}{\cal Z}_{\g_h}] ={\rm Im}[\zeta^{-1}{ Z(\g_h)}]
-\sum_{A'} \frac{\langle \gamma_h,\gamma_{A'}\rangle/2}{|\,\vec x-\vec x_{A'}\,|}\ .
\end{equation}
This means that the interaction is analogous to that of a unit electrically
charge particle in a magnetic monopole of total magnetic flux
$2\pi\langle \gamma_h,\gamma_{A'}\rangle$ while being also constrained
by a radial potential with minimum at some finite radius.

This simple and universal low energy dynamics of a probe dyon
is a direct consequence of (\ref{BPS}),
which can be trusted whenever one is near a marginal stability wall.
Note how this probe-dynamics is
entirely determined by the single central charge function $\CZ_{\gamma_h}(\vec x)$.
A little further away from
such a wall, one actually find the potential to be
\begin{equation}
|\CZ_h|-{\rm Re}[\zeta^{-1}\CZ_h]\ ,
\end{equation}
but this reduces to
\begin{equation}
\CK^2/2|\CZ_h|=({\rm Im}[\zeta^{-1}\CZ_h])^2/2|\CZ_h|\ ,
\end{equation}
as we move near
the marginal stability wall where phases of $Z(\g_h)$ and $Z(\g_c=\sum_{A'}\g_{A'})$
line up \cite{Lee:2011ph}. This approximation is more than good enough
since we already know that BPS spectrum are continuous far away
from marginal stability walls; for the purpose of BPS state counting,
we are allowed to go as close to an MSW as we wish, as long as we do
not cross it.

When we treat all charge centers on equal footing, the low
energy dynamics can be quite complicated. The part least affected
by this is the Lorentz force, thanks to its topological nature.
The half-integral Dirac-quantized coefficient in $\CW$ encodes
how one particle's electric (magnetic) charge see the other particles'
quantized magnetic (electric) charge. Because of the Bianchi
identity the basic shape of $\CW$ cannot be corrected either,
and also such interactions arise only from sum of two-body
interactions. Therefore, these one-derivative interaction
terms are computed by adding up all pair-wise topological couplings,
as
\begin{equation}
-\frac{d\vec x}{dt} \cdot\vec \CW \quad\rightarrow \quad
-\frac{d\vec x_A}{dt} \cdot\vec \CW_A \ ,
\end{equation}
where \cite{Kim:2011sc}
\begin{equation}\label{WA}
\vec\CW_{A} = \frac12 \sum_{B\neq A} \langle \gamma_A,\gamma_{B}
\rangle\,\vec\CW^{Dirac}(\vec x_A-\vec x_B) \ .
\end{equation}
$\CW^{Dirac}$ is the Wu-Yang vector potential \cite{Wu:1976ge}
of a $4\pi$  flux Dirac monopole.

It turns out that this also fixes the scalar potential
completely; see section 5 for more detail. The answer is
\begin{equation}
\frac{{\cal K}_h^2}{2f} \quad\rightarrow \quad
\frac12(f^{-1})^{AB} {\cal K}_A{\cal K}_B
\end{equation}
with \cite{Kim:2011sc}
\begin{equation}\label{KA}
{\cal K}_A={\rm Im}\left[\zeta^{-1}_T{\cal Z}_A\right]={\rm
Im}\left[\zeta^{-1}_T{Z}(\g_A)\right]
-\frac12\sum_{B\neq A}\frac{\langle\g_A,\g_B\rangle}{|\,\vec x_A-\vec x_B|}\ .
\end{equation}
Here, $f_{AB}$ is $(n+1)$-particle version of the
inertia function $f$, i.e.,
\begin{equation}
\frac12\, f \left(\frac{d\vec x}{dt}\right)^2 \quad\rightarrow \quad
\frac12\, f_{AB}\, \frac{d\vec x_A}{dt}\cdot\frac{d\vec x_B}{dt} \ .
\end{equation}
$f_{AB}$ cannot be computed precisely with the above approach;
one would need to repeat the probe-core dynamics with the core dyons slowing
moving, in order to derive these kinetic functions in their full
generality. Fortunately for us, details of $f_{AB}$
is not important as long as zero locus of ${\cal K}_A$
carves out a large manifold far away from the ``origin,"
$\vec x_A=\vec x_B$. See section 6. For most field theory BPS states,
this limit suffices for the wall-crossing physics.

\section{Subtlety: $3n$ vs. $2n$}

The low energy dynamics of such $(n+1)$ dyons include a scalar
potential, $(f^{-1})^{AB} {\cal K}_A{\cal K}_B/2$, so it
sounds natural to reduce it further to the classical
moduli space, ${\cal K}_A=0$. Of $3n+3$
position coordinates, three center of mass coordinates would
decouple, so the reduction would localize the relative part of
the dynamics with $3n$ coordinates to $2n$ coordinates. It would seem
that, for the purpose of finding ground states,
one  have an option to simplify the problem as,
\begin{equation}
\mathbb{R}^{3n+3} \qquad\Rightarrow \qquad \mathbb{R}^3_{\rm c.o.m.}\times {\cal M}_{n+1}\ ,
\end{equation}
where ${\cal M}_{n+1}\equiv \{\vec x_1,\vec x_2,\dots, \vec x_{n+1}\;
\vert\; {\cal K}_A=0\}/\mathbb{R}^3_{\rm c.o.m.}$.
Among other things, this  would immediately imply that
supersymmetric ground state exists only if ${\cal M}_{n+1}$ is
not empty. The latter require the signs of ${\rm Im}\left[\zeta^{-1}_T{Z}_A\right]$
be appropriately correlated with those of $ \langle \g_A,\g_B\rangle $.
MSW's are located where one or more of ${\rm Im}\left[\zeta^{-1}_T{Z}(\g_A)\right]$'s vanish.

However, this naive truncation is unjustified at
quantum level, and cannot be taken at the face value.
Before seeing why this is the case, it is worthwhile to note how
it had  failed in the past.
Consider a two-particle case. As mentioned in the first section,
this problem has been addressed by many different approaches
including that of Denef \cite{Denef:2002ru}, and the correct answer for the number
of ground states is $\vert\langle \g_1,\g_2\rangle\,\vert$ times
individual degeneracies of the two constituent particles.
On the other hand, ${\cal M}_2$ is  a two-sphere threaded by
the magnetic flux $2\pi\langle \g_1,\g_2\rangle$. Since one is
supposed to find BPS states that belong to $D=4$ $N=2$ theory,
she should expect four (real) unbroken  supercharges control the low
energy dynamics. A two-sphere is not a hyperK\"ahler manifold, so
the best we can expect for the reduced dynamics with four real
supercharges is a nonlinear sigma model onto ${\cal M}_2=S^2$ with
${\cal N}_C=2$ complex supersymmetry. However, this is a well-known
quantum mechanics problem, whose ground state counting gives
$\vert\langle \g_1,\g_2\rangle\vert+1$, instead. The correct
answer $\vert\langle \g_1,\g_2\rangle\vert$, on the other hand, is
known to arise if one assumes smaller supersymmetry, which generates even more questions.

The reason for the failure can  be understood easily
with the same two particle example. Note that, near ${\cal M}_2$,
the massive radial mode $\delta r\equiv |\vec x_1-\vec x_2|-R_{12}$ enters the potential as
\begin{equation}\label{HO}
\frac{{\cal K}^2}{2f}\simeq
\frac{1}{2f(R_{12})}\left(\frac{\langle \g_1,\g_2\rangle/\,2}{(R_{12})^2}\right)^2\delta r^2\ ,
\end{equation}
where $R_{12}\equiv \langle\g_1,\g_2\rangle/2{\rm Im}\left[\zeta^{-1}_T{Z}_1\right]
=\langle\g_2,\g_1\rangle/2{\rm
Im}\left[\zeta^{-1}_T{Z}_2\right]$. The other, angular directions
appear massless, and thus are deemed to be lower energy degrees of
freedom.

This reasoning would have held if we were either dealing with
classical mechanics or with higher dimensional quantum field theory. For
quantum mechanics, a massgap can arise not only from massive potential
but also when the target is of finite volume. A canonical example
is the one-dimensional potential well of width $l$, whose
lowest energy eigenvalue goes like $\sim 1/l^2$. Here the classical
moduli space is $S^2$ of radius $R_{12}$, and the angular momentum
part of the wavefunction contributes to the quantum energy,
\begin{equation}
\frac{1}{2f(R_{12})}\cdot \frac{\vec L\cdot\vec L-(\langle \g_1,\g_2\rangle/\, 2)^2}{(R_{12})^2}
\end{equation}
with the angular momentum operator $\vec L$. Because of the magnetic
flux in the background, both $\vec L$ gets shifted by the amount
$\langle \g_1,\g_2\rangle/2 \times(\vec x_1-\vec x_2)/|\vec x_1-\vec x_2|$, and
the lowest eigenvalue for $\vec L \cdot\vec L$ is $|\langle \g_1,\g_2\rangle/2|^2+
|\langle \g_1,\g_2\rangle/2|$.
The massgap associated with these classically massless directions are thus
\begin{equation}
\frac{|\langle \g_1,\g_2\rangle/\, 2|}{2f(R_{12}) \times (R_{12})^2}\ .
\end{equation}
Note that this massgap equals exactly
the ground state energy of the $\delta r$ harmonic oscillator
(\ref{HO}) with $f(R_{12})$ being the  inertia for $\delta r$.
In short \cite{Kim:2011sc}, at quantum level, there is no further natural reduction
of relative dynamics on $\mathbb{R}^{3}$ to the classical moduli space ${\cal M}_2=S^2$.

One can easily trace this equality among massgaps,
to the identity
\begin{equation}
d{\cal W}= *d{\cal K}\ ,
\end{equation}
which, as we will see below, is due to ${\cal N}=4$ supersymmetry of the
low energy quantum mechanics. As was already stated in (\ref{WA}) and
(\ref{KA}), such a close relationship between the scalar potential
and the vector potential is not limited to two particle problem
but holds true for arbitrary $(n+1)$ particle  problem. There again,
$\CN=4$ supersymmetry demands it. Therefore, {\it despite the
very useful  classical picture of charge centers at fixed
mutual distances, one is not allowed to  formulate the low energy
dynamics on the classical moduli space, ${\cal M}_{n+1}$, spanned by
such solutions.}

\section{$\CN=4$ Supersymmetry}

Supersymmetrization of the low energy dynamics can be performed most
economically in the ${\cal N}=1$ superspace notation. This choice
is also convenient because, later, we end up mathematically deforming the
dynamics to an $\CN=1$ nonlinear sigma model for facilitating index
computation.

Denote by $\vec x^{A}$
the position vector  of the charge $\g_A$  center in the real space
$\mathbb{R}^3$. The collective coordinate degrees of freedom
for each charge center can be put into four $\CN=1$ superfields,
\begin{equation}
\Phi^{Aa}=x^{Aa}-i\theta\psi^{Aa}\ ,\quad
\Lambda^A=i\lambda^A+i\theta b^A \ ,
\end{equation}
with an auxiliary field $b^A$ introduced for the fermionic superfield $\Lambda^A$.
The Lagrangian that
supersymmetrizes the low energy dynamics of section 3
can be written as \cite{Lee:2011ph,Kim:2011sc}
\begin{eqnarray}
 {\cal L}&={\cal L}_0+{\cal L}_1\ ,
\end{eqnarray}
where
\begin{eqnarray}
{\cal L}_1&=& \int d\theta\; \left(i{\cal K}_A(\Phi)\Lambda^A -i{\cal
W}_{Aa}(\Phi) D\Phi^{Aa}\right) \ , \label{L1many}
\end{eqnarray}
encodes the scalar and the vector potentials;
the kinetic term ${\cal L}_0$ will be discussed later.
One can show
\begin{equation}
\delta_\epsilon\int dt\; {\cal L}_1=0
\end{equation}
under ${\cal N}=4$ supersymmetry transformation rules,
with $\psi^{A4}\equiv\lambda^A$,
\begin{eqnarray}\label{SUSY}
\d_\e x^{A}&=& i \eta^{a}_{mn} \e^m \psi^{An} \ , \qquad
\nonumber\\
\d_\e \psi_m^A &=& \eta^a_{mn} \e^n {\dot x}^{Aa} + \e_m b^A\ ,
\nonumber \\
  \d_\e b^A &=& -i\e_m\dot \psi^{Am}\ ,
\end{eqnarray}
 provided that
\begin{eqnarray}\label{n=4}
\partial_{Aa}{\cal
K}_B&=&\frac12\,\epsilon_{abc}\left(\partial_{Ab}{\cal
W}_{Bc}-\partial_{Bc}{\cal W}_{Ab}\right)\ ,
\end{eqnarray}
 and
\begin{eqnarray}\label{n=4'}
\epsilon_{abc}\partial_{Ab}\partial_{Bc}\CK_C&=&0\ ,\nonumber\\
\partial_{Aa}\partial_{Ba}\CK_C&=&0\ .
\end{eqnarray}
If these constraints are not met, $\CL_1$ would be invariant
under the one manifest supersymmetry, which corresponds to $\epsilon^4$ in (\ref{SUSY}).

That these $\CN=4$ constraints are solved by (\ref{WA}) and (\ref{KA})
should be obvious to readers. As noted already in section 3,
${\cal W}_A$ are of topological nature and cannot be corrected
by higher order effects. When obtaining ${\cal K}_A$'s from $\CW_A$'s via
(\ref{n=4}) and (\ref{n=4'}), the only extra input needed is the asymptotic values,
$\CK_A(\infty)={\rm Im}\left[\zeta^{-1}_T{Z}_A\right]$. However,
these are tied to the energy cost when we separate $\g_A$ dyon
center to spatial infinity and is determined unambiguously from
the superalgebra. Therefore, despite the fact that we are dealing
with dyons in strongly coupled theories, the interaction Lagrangian
with one or less time derivative, ${\cal L}_1$, is fixed
without any ambiguity at all. The small control parameters are
the inverse of the classical distances between centers, which are
in turn held small by the proximity to MSW's.

The kinetic term $\CL_0$ is a little more involved.
The simplest way to find general form of $\CL_0$ is
to note that the collection $\{\Phi^{A1},\Phi^{A2},\Phi^{A3},\Lambda^A\}$
for each $A$ can be thought of as dimensional reduction
of a $D=4$ $N=1$ vector superfield \cite{Ivanov:1990jn,Berezovoj:1991ka}.
In the Wess-Zumino gauge of the latter,
$x^a$'s come from the spatial part of the vector gauge field,
the fermions from the gaugino, and the auxiliary field $b$ from
the $D=4$ auxiliary field. As such, $N=1$
the superspace of the latter can be used as $\CN=4$
superspace here. Let us again display $\CN=1$ form of
such a general $\CN=4$  $\CL_0$ as, \cite{Maloney:1999dv}
\begin{eqnarray}
 {\cal L}_0&=&\int  d\theta \;\frac{i}{2}\,g_{AaBb}
D\Phi^{Aa}\partial_t\Phi^{Bb} -\frac12\, h_{AB}\Lambda^A
D\Lambda^B
-ik_{AaB}\dot\Phi^{Aa}\Lambda^B +\cdots \ .
\end{eqnarray}
where the ellipsis denotes four cubic terms that we omit here
for the sake of simplicity. ${\cal L}_0$ is also invariant
under the four supersymmetries we listed above,
\begin{equation}
\delta_\e\int dt \,{\cal L}_0=0
\end{equation}
with off-shell $b^A$'s, provided that
various coefficient functions derive from a single real function
$L(x)$ of $3n+3$ variables as
\begin{eqnarray}
g_{AaBb}(\Phi)&=&\left(\delta^e_a\delta_b^f+\e^{\;\;e}_{c\;\;a}\e^{cf}_{\;\;\;
b}\right)
\partial_{Ae}\partial_{Bf}L(\Phi) \ , \nonumber\\
h_{AB}(\Phi)&=&\delta^{ab}\partial_{Aa}\partial_{Bb}L(\Phi)\ ,\nonumber\\
k_{AaB}(\Phi)&=&\e^{ef}_{\;\;\;a}\partial_{Ae}\partial_{Bf}L(\Phi)\ ,
\nonumber\\
&\vdots&
\end{eqnarray}
Figuring out the precise form of the function $L$ for $n+1$ charge centers
requires further work. However, for the purpose of deriving wall-crossing
formula for field theory BPS states, the asymptotic form of $L$ should
suffice, as we argue below. In this limit, $L$ encodes only the masses of
individual charge centers as
\begin{equation}
L\;\simeq\; -\frac12\sum_A |Z(\g_A)|\,\vec  x_A\cdot \vec x_A\ .
\end{equation}

\section{Deformation to ${\cal M}_{n+1}$ and Index Theorem}

Now that we supersymmetrized the low energy dynamics, the discussion of
section 5 extends easily to fermions as well. Of four fermions for each $A$, one pair
acquires mass from the bilinear coupling to $d\CK_A$ while the other pair
become massive via such a coupling to $d\CW_A$. With the tight constraint
between $d\CK_A$'s and $d\CW_A$'s, it is clear that the fermions cannot
be divided into massive ``radial" and massless ``angular" modes, either.
In principle, one could proceed with the above Lagrangian and compute
relevant indices in $\mathbb{R}^{3n}$. However, it turns out that the dichotomy
between classically massive and classically massless direction can be
salvaged yet, simplifying the index computations and bringing us to the
classical moduli space ${\cal M}_{n+1}$ after all.

The key observation is that, in defining a supersymmetric index (i.e. the
difference between the number of bosonic and the number of fermionic
ground states), we
only need one supercharge and one chirality operator.
Although the full dynamics has $\CN=4$ supersymmetry, only
one supercharge is needed for the computation. Furthermore, supersymmetric
index is a topological quantity and insensitive to  ``small"
deformations of dynamics. As we formulated the low energy dynamics of
dyons in terms of $\CN=1$ superspace, we may as well keep the one
manifest supersymmetry and deform the dynamics as we see fit, without
affecting the index at all.
Therefore, the same index can be computed
from a different dynamical system, say, the one with the potential
part of Lagrangian given as \cite{Kim:2011sc}
\begin{eqnarray}\label{deform}
{\cal L}_1^\xi &=& \int d\theta\; \left(i\xi {\cal K}_A(\Phi)\Lambda^A -i{\cal
W}_{Aa}(\Phi) D\Phi^{Aa}\right) \ , \label{L1many}
\end{eqnarray}
for an arbitrary real positive number $\xi$. This clearly breaks $\CN=4$
down to $\CN=1$ but has the advantage of making the ``radial" direction
far heavier than angular directions.

The equality of massgaps between
the two sectors is disrupted by $\xi$; radial massgaps are now $\xi$ times
larger than the angular massgaps.
With $\xi\rightarrow \infty$, then, $n$ radial modes become infinitely
heavy and can be decoupled from $2n$ modes along $\CM_{n+1}$.
Similarly the $2n$ fermions coupled $\xi d \CK$'s
will become infinitely heavy, leaving behind the other $2n$ coupled to  $d\CW$'s.
It has been shown rigorously \cite{Kim:2011sc} that this deformation leaves
behind an $\CN=1$ nonlinear sigma model with the Lagrangian
onto $\CM_{n+1}$
\begin{equation}
{\cal L}_{\rm for \; index\; only}^{\CN=1}({\CM_{n+1}})=\frac12\,
g_{\mu\nu}\dot z^\mu\dot z^\nu
+\frac{i}2 \,g_{\mu\nu} \psi^\mu\dot \psi^\nu+\cdots-{\cal
A}_\mu\dot z^\mu+\cdots\ ,
\end{equation}
schematically, where $\CA$ is an external Abelian gauge field on $\CM_{n+1}$,
\begin{equation}
d\CA={\cal F}\equiv d\left(\sum_A
\CW_{Aa}dx^{Aa}\right)\Biggl\vert_{\, \CM_{n+1}}\
\end{equation}
and the induced metric $g$ on $\CM_{n+1}$ from the ambient $\mathbb{R}^{3n}$.

It is important to remember that the reduction applies to the relative
part of the dynamics.
The free center of mass part $\mathbb{R}^3$  remains intact with its own $\CN=4$
supersymmetry, and supplies the obligatory half-hypermultiplet structure
to bound states that may emerge from the relative dynamics.
Among other things, this also implies that the second helicity trace
is computed in the relative dynamics as the usual Witten index
with $(-1)^{2J_3}$ as the chirality operator, whose precise
definition in the quantum mechanics need a bit more of clarification.
Anyhow, since the surviving supersymmetry clearly reduces to the real
supersymmetry on $\CM_{n+1}$ twisted by the magnetic field $\CF$,
and since the fermionic partners are real, we come to the Dirac index,
\begin{equation}
{\rm Tr}\left((-1)^{F_{\CM_{n+1}}}e^{-\beta Q^2}\right)
=\int_{{\cal M}_{n+1}}Ch({\cal F})\hat A({\cal M}_{n+1})\ ,
\nonumber
\end{equation}
with the Chern character $Ch$ and A-roof genus $\hat A$.
This index theorem is the most basic ingredient that will
eventually compute $\Omega\left(\sum_A\g_A\right)$, although
there are more things to do before we can connect to actual
bound state spectrum and the wall-crossing formulae.

The first thing to clear up is the chirality
operator $(-1)^{F_{\CM_{n+1}}}$, which in this equation is
defined so that we have the canonical index formula. What is not
immediate, however, is how this chirality operator is related to
$(-1)^{2J_3}$. The answer to this has the universal form, \cite{Kim:2011sc}
\begin{equation}
(-1)^{2J_3}\quad\rightarrow\quad
(-1)^{\sum_{A>B}\langle\g_A,\g_B\rangle +n}(-1)^{F_{\CM_{n+1}}}
\end{equation}
bringing us to
\begin{equation}
\CI_{n+1}(\{\g_A\})\equiv
(-1)^{\sum_{A>B}\langle\g_A,\g_B\rangle +n}
\int_{{\cal M}_{n+1}}Ch({\cal F})\hat A({\cal M}_{n+1})\ .
\end{equation}
as the right index theorem.

Second, as one approaches an MSW, the zero locus of the potential, $\CK_A=0$,
would expand to the asymptotic region,
where the ambient space $\mathbb{R}^{3n}$ near such
large $\CM_{n+1}$ would be flat for all intent and purpose.
Recall that there is no discontinuity before one reaches MSW; spectrum
should be independent of how close we get to the MSW, or equivalently
indendent of how large $\CM_{n+1}$ becomes as long as the latter remains finite.
With $g$ induced from the
flat $\mathbb{R}^{3n}$, ${\cal M}_{n+1}$ turns out  to  carry a trivial
A-roof genus and the above collapses to a symplectic volume, \cite{Kim:2011sc}
\begin{equation}\label{symplectic}
\CI_{n+1}(\{\g_A\})=
(-1)^{\sum_{A>B}\langle\g_A,\g_B\rangle +n}
\int_{{\cal M}_{n+1}}e^{{\cal F}/2\pi}\ ,
\end{equation}
simplifying the problem of evaluation enormously.
For two particle problems, in particular, this translates to
\begin{equation}
\pm\, \langle\g_1,\g_2\rangle\ ,
\end{equation}
which, as we mentioned already, is the right index for
primitive wall-crossing. In particular, this without
the wrong $+1$ shift that would have resulted from naive
truncation of section 3.

Third,  we should take care to include the degeneracies,
or the indices $\Omega(\g_A)$, of individual charge centers as well.
In the above derivation, an implicit assumption was that such
internal degeneracies of constituent particles do not interfere with
the bound state formation. As such, they will simply contributes
a multiplicative factor to the bound state counting.
The index $\Omega^-$, counted from the second helicity
trace introduced in section 1, should be computed from $\CI$ as
\begin{equation}
\Omega^-\left(\sum_A\g_A\right)=\CI_{n+1}(\{\g_A\})\prod\Omega(\g_A) \ .
\end{equation}
$\Omega(\g_A)$ has no $\pm$ superscript here since by construction
we are considering $\g_A$'s that are point-like on both sides of
MSW; such states would have the same intrinsic index on the two
sides of the wall, since wall-crossing affects loose bound
states. By the way,
it could happen, in principle, that $+$ side has states of total
charge $\sum_A\g_A$ but of different origin. In such cases, the
left hand side is to be understood as $\Omega^- -\Omega^+$.

The final issue, which is of more fundamental nature, comes
from the fact that we treated $\g_A$'s as all distinguishable.
In most wall-crossing problems, however,
we end up counting bound states of type $n\g_1+k\g_2$
for positive integers $n$ and $k$. Statistics is thus a vital
issue. In next section, we will see how this gets incorporated
into the problem, whereby  the rational invariants seen in the
Kontsevich-Soibelman formalism of section 2 will re-emerge.

\section{Statistics, Rational Invariants, and Wall-Crossing Formula}

Statistics can be imposed on top of the index problem by
inserting a projection operator that symmetrizes or anti-symmetrizes
wavefunctions under permutations of identical particles. Given each
permutation group $S(k_i)$ of $k_i$ identical particles, we need to insert
the projection operator \cite{Kim:2011sc}
\begin{equation}
\CP(k_i)=\frac{1}{k_i!}\sum_{\sigma\in S(k_i)} (\pm 1)^{|\sigma|}\sigma
\end{equation}
for bosonic and fermionic constituent particles, respectively.
As we expect typically more than one species of particles involved,
we should insert such projections for each  species. On
most of configuration space, this projection will act freely
and the net result would be a division of the target volume
by $1/k_i!$.

However, there are fixed submanifolds spanned by
configurations where, say, $k_i^1$ identical particles are moving
together. This generates an additive contribution. Carving out a tubular neighborhood of
such submanifolds, we can see that the dynamics along this submanifold
will look exactly as before except that, instead of $k_i$ identical
particles of charge $\g_i$, we see $(k_i-k_i^1)$ identical particles of
charge $\g_i$ plus a single additional particle of charge $k_i^1\g_i$.
For the extra contribution from such fixed submanifold,  we must now
insert a projection operator associated with $S(k_i-k_i^1)=S(k_i)/S(k_i^1)$
permutation group, and repeat the exercise. From this general discussion,
it is clear that the index with the projector
$\CP_i$'s inserted can be iteratively decomposed to many index problems,
each of which has the same total charge but smaller number of charge centers.

The projected index theorem goes as, with the two representative
terms discussed above shown explicitly,
\begin{eqnarray}
\CI_{\CP}\left(\sum_i k_i \g_i\right)
&=&{\rm Tr}\left((-1)^{F_{\CM_{n+1}}}e^{-\beta Q^2}\prod\CP(k_i)\right)\cr\cr
&=&\frac{1}{\prod_i k_i!}\,
\CI_{n+1}(\{\,\cdots,\g_i,\g_i,\cdots,\g_i, \g_i,\cdots\}\,)\cr\cr
&&\qquad \vdots\cr
&&+\frac{\Delta(k^1_i)}{(k_i-k^1_i)!\prod_{j\neq i}k_j!}\,
\CI_{n+1-k^1_i+1}(\{\, \cdots,k_i^1\g_i,\g_i,\cdots,\g_i,\cdots   \}\,)\cr\cr
&&\qquad\vdots
\end{eqnarray}
where each and every $\CI$ computes bound states of
the same total charge and
\begin{equation}\label{D}
\Delta(p)=\pm 1/p^2
\end{equation}
is a universal factor associated with $p$ coincident
identical (bosonic/fermionic) particles. The origin of $\Delta$ will be
explained below. Once we accept the latter, it should be clear how
terms in the above sum are generally constructed.
For each and every partition
\begin{equation}
k_i=\sum_\alpha k^\alpha_i\ , \qquad k_i^\alpha\ge 1 \ ,
\end{equation}
treat $k_i^\alpha\g_i$ as if it
is an individual particle and compute $\CI$ for this reduced
dynamics with less number of particles. Then the additive
contribution to $\CI_\CP$ from such a  partition is
\begin{equation}
\frac{\prod_{i,\alpha}\Delta(k_i^\alpha)}{\prod_{i}
\left\vert S(k_i)/(S(k^1_i)\times S(k_i^2)\times\cdots)\right|}
\,\CI(\{k_1^1\g_1,k_1^2\g_1,\cdots\}) \ .
\end{equation}
The denominator is clearly the volume-reducing factor from
the residual permutation symmetry, and we again have the numerical
factors $\Delta(k_i^\alpha)$ for each and every coincident
identical particles. $\CI_\CP\left(\sum_i k_i \g_i\right)$ is a sum of such terms over
all possible partitions of $\{k_i\}$.

For $\CI_\CP$, then, it remains to derive the multiplicative
factors $\Delta(k_i^\alpha)$. In the above, we evaluated the projected
index $\CI_\CP$ by decomposing it according to how the
permutation groups act. Each additive contribution arise
from carving out an infinitesimally thin tubular neighborhood
around a fixed submanifold where $k_i^\alpha$ of $\g_i$ charge
centers coincides and move together. In this
tubular neighborhood, however, there are also directions
associated with separating $k_i^\alpha\g_i$
into $k_i^\alpha$ number of $\g_i$'s. They are the fibre directions
of the normal bundle in $\CM$ of such a fixed submanifold.
On top of $\CI(\{k_1^1\g_1,k_1^2\g_1,\cdots\})$,
then, one should expect to see a multiplicative factor associated
with these directions.

Since these $k_i^\alpha$ $\g_i$'s are identical and of the same charge,
there are no interaction of type $\CL_1$; Therefore, each $\Delta(k_i^\alpha)$
should be computed from free $k_i^\alpha$ particle dynamics, except that
the projector $\CP(k_i^\alpha)$ should be inserted to correctly take
into account the statistics.
 One might think that, since free dynamics cannot lead to
a bound state, so the corresponding factor $\Delta(k_i^\alpha)$ should
vanish. However, when computing the above index $\CI_\CP$, we are
actually computing the so-called bulk term which captures continuum
contributions as well; for the full index this is good enough
because $\CM_{n+1}$ is a compact manifold and for $\Delta(k_i^\alpha)$
computation, the same limit should be used consistently since this
part of computation is embedded in the index computation on the same
compact $\CM_{n+1}$.

Interestingly, exactly such a quantity $\Delta(p)$ was computed
some fifteen years ago, in the context of supersymmetric Yang-Mills
quantum mechanics \cite{Yi:1997eg,Green:1997tn}. The problem back then was
whether or not identical (and wrapped) D-branes can have a threshold bound
state; a particular case of this with the maximal supersymmetry is
the famous D0 bound state problem \cite{Witten:1995im}. $\Delta(p)$
appeared there as a defect term, coming from a continuum contribution
from $p$ identical and unbound particles,
and had the universal form
\begin{equation}
\Delta^{YMQM}(p)=+\frac{1}{p^2}\ ,
\end{equation}
for any allowed supersymmetry. The $+$ sign shows up because the bosonic
statistics is built-in for the Yang-Mills quantum mechanics, via
the Weyl group. Repeating the exercises with the fermionic statistics
allowed, we finds (\ref{D}).

We must go one step further and figure out
how the  indices $\Omega(\g_i)$'s of the constituent particles
affects the computation of $\Delta$'s, which lead to \cite{Kim:2011sc}
\begin{eqnarray}
\Omega^-\left(\sum_i k_i\g_i\right)
&=&\frac{\prod_i \Omega(\g_i)^{k_i}}{\prod_i |S(k_i)|}\times \CI\,(\{\g_1,\g_1,\cdots\}\,)\cr\cr
&&\qquad\vdots\cr
&&+\frac{\prod_{i,\alpha}\Omega(\g_i)/(k_i^\alpha)^2}{\prod_{i}
\left\vert S(k_i)/(S(k^1_i)\times S(k_i^2)\times\cdots)\right|}
\times \CI\,(\{k_1^1\g_1,k_1^2\g_1,\cdots\}\,)\cr\cr
 &&\qquad\vdots \
\end{eqnarray}
where one last ingredient we used is that the positive and the negative
$\Omega(\g_i)$ imply, respectively, the fermionic and the bosonic statistics
for $\g_i$. This computes the bound state index of charge $\sum_i k_i\g_i$
made from $\sum_ik_i$ such charge centers.

For the most general wall-crossing formula, we may allow
the logical possibility that such a BPS state can be built differently,
when other constituent charge centers are available, say,
$\Omega^+(2\g_1)\neq 0$ or $\Omega^+(2\g_1+3\g_2)\neq 0$,
etc. Adding up all such contributions yet again, we find that
the final answer is a sum
over all possible partition, $\{\g_K\}$,
\begin{equation}
\g_T\equiv\sum_A\g_A=\sum_K \g_K \ ,
\end{equation}
where $\g_A$, some of which can be indistinguishable, are all primitive. By
a partition, we mean that each $\g_K$ is a nonnegative integral
linear combination of $\g_A$'s. Each partition generates an additive
contribution so that the final index on the - side of MSW is \cite{Kim:2011sc,Manschot:2010qz}
\begin{eqnarray}
\Omega^-\left(\sum_A\g_A\right)\quad=\quad\cdots\quad
+\frac{\prod_{K}\bar \Omega(\g_K)}{|S(\{\g_K\}\,)|}
\times \CI\,\left(\{\g_K\}\,\right)+\quad \cdots
\end{eqnarray}
with the residual permutation group, $S(\{\g_K\})$, and the
rational invariants,
\begin{equation}
\bar\Omega(\g_K)=\sum_{s\vert \g_K} \frac{\Omega(\g_K/s)}{s^2} \ .
\end{equation}
To be more precise, one must take care to keep track of
flavor charges in $\g$'s as well, to avoid potential
ambiguities in this formula.

\section{Protected Spin Character and Equivariant Index}

A more general index that keeps track of global quantum numbers
of states, beyond counting degeneracies, is
known and dubbed the protected spin character (PSC) \cite{Gaiotto:2010be},
\begin{equation}
\Omega_{\rm PSC}(y)=-\frac12\,{\rm tr}
\left((-1)^{2J_3}(2J_3)^2y^{2J_3+2I_3}\right)\ . 
\end{equation}
$J_3$ and $I_3$ are generators of, respectively, the little group
$SU(2)_L$ and the R-symmetry $SU(2)_R$. Of $SU(2)_R\times U(1)_R$ R-symmetry,
the latter factor is ``spontaneously broken" by any given BPS state, as
the central charge phase rotates under $U(1)_R$.
How such an equivariant generalization descends to the low energy
quantum mechanics deserves a brief explanation, as it also have caused some
confusion in the past.

For the true low energy dynamics involving $3n+3$
bosonic coordinates, the descent is actually straightforward. Bosons $x^{Aa}$'s
and fermions $\psi^{Am}$ transform naturally under $SO(4)=SU(2)_+\times SU(2)_-$,
as (3,1) and (2,2), respectively. Since $x^{Aa}$'s encode the positions
of charge centers, they rotate as vectors under $SU(2)_L$ but must be
invariant under $SU(2)_R$. From this it is then
clear that $SU(2)_+=SU(2)_L$ and $SU(2)_-=SU(2)_R$.
As such, PSC descend to a quantum mechanical equivariant index,
\begin{equation}
\Omega_{\rm PSC}(y)\quad\rightarrow\quad
\Omega(y)={\rm tr}\left((-1)^{2J_3}y^{2J_3+2I_3}\right)\ , 
\end{equation}
verbatim, with $J_3$ and $I_3$ now understood to be those of the low energy dynamics.
We further factored out the free center of mass part of Hilbert space,
which effectively drops $-(2J_3)^2/2$ and instead traces only over the
relative part of the dynamics. When $y=1$, this is precisely the index
we computed above.

In deforming the dynamics for index computation,
down to the nonlinear sigma model on $\CM_{n+1}$, one
must be a little more careful, as the process does not preserve $SO(4)$
global symmetry. Again this can be illustrated easily with the minimal example
of two particle system, where $\CM_2=S^2$. The latter manifold admits
only one isometry group, call it $SO(3)_\CJ$, whose origin in
$SO(4)=SU(2)_L\times SU(2)_R$ should be clarified. Recall
that, after the deformation and taking the limit $\xi\rightarrow \infty$,
one finds a nonlinear sigma model onto $\CM_2=S^2$.
In nonlinear sigma models, bosonic fluctuations $\delta z^\mu$ and
fermionic fluctuations $\psi^\mu$ transform in the same manner under
coordinate transformations and also under isometry. This means that
$SO(3)_\CJ$ transform the surviving $2n$ bosonic angles and $2n$
fermionic partners by the same rule. Each originate from (3,1) and
(2,2) of $SO(4)$, and clearly this implies that the surviving isometry
is the diagonal subgroup. That is,
\begin{equation}
\CJ_a= J_a+I_a  \ .
\end{equation}
The same is easily seen to be true of general $(n+1)$ particle problems.
Furthermore, it is obvious from (\ref{SUSY}) that this $\CJ$
commute with the single manifest $\CN=1$ supersymmetry, associated with $\epsilon^4$,
of $\CL_1^\xi$ in (\ref{deform}). The latter is also what becomes
the supersymmetry of $\CM_{n+1}$ nonlinear sigma model.

Therefore, PSC of the field theory reduces to the $\CM_{n+1}$ index as
\begin{eqnarray}
\Omega_{\rm PSC}(y)\qquad\rightarrow\qquad
\Omega(y)&=&{\rm tr}_{\CM_{n+1}}\left((-1)^{2J_3}y^{2\CJ_3}\right) \ ,
\end{eqnarray}
where $(-1)^{2J_3}$ still makes sense as a chirality operator, even
though $SU(2)_L$ is broken by $\xi\neq 1$. (See section 6 for more
on this chirality operator.) This is the usual equivariant index for the
nonlinear sigma model, up to an overall sign of the chirality operator
that we took care to fix in section 6,
so we finally matched PSC of field theory to equivariant index of $\CM_{n+1}$.
Extension of wall-crossing formula of previous section to such an
equivariant version is straightforward and well-established in
mathematics literatures. We will refer readers
to Maschots et.al. \cite{Manschot:2010qz,Manschot:2011xc}
for detailed exposition on these equivariant quantities
in the current context, as well as
for explicit evaluations.

\section{Conclusion and Beyond Multi-Center Picture}

In this talk, we reviewed how the multi-center picture of BPS dyons
leads to an intuitive understanding of wall-crossing,
and outlined how one derives low energy dynamics for such
semiclassical objects even in strongly coupled regime.
We then derived general wall-crossing formula for $D=4$
$N=2$ supersymmetric field theories. Along the process, we
clarified when the low energy dynamics may be used, how
it should be formulated as $\mathbb{R}^{3n+3}$ quantum
mechanics rather than $\CM_{n+1}$ nonlinear sigma model,
how the field theory index descends to those of the low
energy dynamics, and finally why the Dirac index on $\CM_{n+1}$
is the relevant one despite the wrong supersymmetry
and the wrong dynamics.  Bose/Fermi statistics of the constituent
particles are shown to be incorporated in the wall-crossing
formula via rational invariants of Kontsevich-Soibelman, which
was used later to show equivalence of the latter's proposal
and our physically derived one \cite{Sen:2011aa}.

Similar low energy dynamics had appeared in the past as the so-called
Coulomb phase picture of the quiver quantum mechanics \cite{Denef:2002ru}.
These quiver theories arise naturally as low energy theories of
D3 branes wrapped on 3-cycles in Calabi-Yau compactified type IIB
theory; through various dualities, they are potentially capable of
capturing dynamics of large classes of BPS states with four preserved
supercharges.
While our starting point is very different from this, one
could regard the low dynamics we found as an alternate derivation,
in the field theory limit, of Coulomb phase dynamics of relevant
quiver quantum mechanics.
Furthermore, the latter part of our analysis applies
to more general Coulomb phase dynamics such as those for black holes:
deformation of $\CN=4$ relative dynamics to $\CN=1$ $\CM_{n+1}$
nonlinear sigma model,  Dirac index on $\CM_{n+1}$
as the basic counting quantity, statistics via rational invariants,
and the resulting wall-crossing formulae are all straightforwardly
applicable to general quiver theories.

Sometimes, however, the multi-center BPS state picture, inherent to
the Coulomb phase description and typical for field theory BPS states, is known
to miss a large class of states in the quiver theory. One early question in
this topic was whether or not the exponential degeneracy of BPS
black hole might be explained from such multi-center pictures,
but it was soon realized that, with a given quiver, one sometimes
finds states of large degeneracy which appear completely missing
in the usual Coulomb phase description.
In fact, it is easy to construct examples where exponentially
large  number of these extra states appear \cite{Denef:2007vg}
in the Higgs phase instead; the degeneracy in the Coulomb phase
in those examples are at most powerlike.

This tells us that there are
more to BPS state counting than wall-crossing phenomena know about;
In the Higgs phase, one finds more comprehensive ground state
space. Furthermore, index counting there is also more straightforward
in that subtleties we encountered in the Coulomb phase are absent.
However, these advantages come at
the cost of losing the simple and intuitive multi-center picture.
Wall-crossing occurs also in the Higgs phase, with exactly the
same discontinuity as in the Coulomb phase;
what has been missing is a way to distinguish,
among the Higgs phase states, the counterparts of the wall-crossing
multi-center Coulomb phase states from those non-wall-crossing states.
Recently, exactly such a method was devised \cite{Lee:2012sc,Bena:2012hf}.
The proposal classifies Higgs phase ground states into two types
with geometrically distinct origins, and identifies one class
as the counterpart of wall-crossing states. The other, expected to be
non-wall-crossing, must be then naturally an invariant of the quiver
quantum mechanics insensitive to continuous change of parameters.
Both of these claims have been tested extensively \cite{Lee:2012na,Manschot:2012rx}.
This new handle will hopefully provide even powerful and versatile methods
to address BPS states, in particular including a large class of BPS
black holes and microstates thereof.

\section*{Acknowledgments}

I would like to thank Seung-Joo Lee, Sungjay Lee, Heeyeon Kim, Jaemo Park,
Zhao-Long Wang for collaborations on various part of this and related works.
I also benefited much from discussions with Jan Manschot, Boris Pioline,
and Ashoke Sen. This work is supported  by the National Research Foundation
of Korea (NRF) funded by the Ministry of Education, Science and Technology
with the grant number 2010-0013526.


\begin{thebibliography}{99}





\bibitem{Prasad:1975kr}
  M.K.~Prasad and C.M.~Sommerfield,
``{An Exact Classical Solution for the 't Hooft Monopole andthe
Julia-Zee
  Dyon},''
  Phys. Rev. Lett.  {\bf 35} (1975) 760.

\bibitem{Bogomolny:1975de}
  E.B.~Bogomolny,
  ``{Stability of Classical Solutions},''
  Sov. J. Nucl. Phys.  {\bf 24} (1976) 449
  [Yad.\ Fiz.\  {\bf 24} (1976) 861].

\bibitem{Seiberg:1994rs}
  N.~Seiberg and E.~Witten,
``{Monopole Condensation, And Confinement In N=2
Supersymmetric Yang-Mills
  Theory},''
  Nucl. Phys.  B {\bf 426} (1994) 19
  [Erratum-ibid.\  B {\bf 430} (1994) 485]
  [arXiv:hep-th/9407087].

\bibitem{Seiberg:1994aj}
  N.~Seiberg and E.~Witten,
``{Monopoles, Duality and Chiral Symmetry Breaking in N=2
Supersymmetric
  QCD},''
  Nucl. Phys. B {\bf 431} (1994) 484
  [arXiv:hep-th/9408099].

\bibitem{Ferrari:1996sv}
  F.~Ferrari and A.~Bilal,
``{The Strong-Coupling Spectrum of the Seiberg-Witten
Theory},''
  Nucl. Phys.  B {\bf 469}, 387 (1996)
  [arXiv:hep-th/9602082].

\bibitem{Bergman:1997yw}
  O.~Bergman,
  ``Three pronged strings and 1/4 BPS states in N=4 superYang-Mills theory,''
  Nucl.\ Phys.\ B {\bf 525} (1998) 104
  [hep-th/9712211].

\bibitem{Lee:1998nv}
  K.M.~Lee and P.~Yi,
``{Dyons in N=4 Supersymmetric Theories and Three Pronged
Strings},''
  Phys. Rev.  {\bf D58 } (1998)  066005.
  [hep-th/9804174].

\bibitem{Bak:1999da}
  D.~Bak, C.K.~Lee, K.M.~Lee, and P. Yi
  ``{Low-energy Dynamics for 1/4 BPS Dyons},''
  Phys. Rev. {\bf D61 } (2000)  025001.
  [hep-th/9906119].


  \bibitem{Gauntlett:1999vc}
  J.P.~Gauntlett, N.~Kim, J.~Park and P. Yi
``{Monopole Dynamics and BPS Dyons N=2 Super Yang-Mills
Theories},''
  Phys. Rev. {\bf D61 } (2000)  125012.
  [hep-th/9912082].

\bibitem{Gauntlett:2000ks}
  J.P.~Gauntlett, C.J.~Kim, K.M.~Lee and P.Yi
``{General Low-energy Dynamics of Supersymmetric
Monopoles},''
  Phys. Rev.  {\bf D63 } (2001)  065020.
  [hep-th/0008031].

\bibitem{Stern:2000ie}
  M.~Stern and P.~Yi,
  ``{Counting Yang-Mills Dyons with Index Theorems},''
  Phys. Rev. {\bf D62 } (2000)  125006.
  [hep-th/0005275].


\bibitem{Denef:2000nb}
  F.~Denef,
  ``{Supergravity Flows and D-brane Stability},''
  JHEP {\bf 0008} (2000) 050
  [arXiv:hep-th/0005049].



\bibitem{Denef:2002ru}
  F.~Denef,
  ``{Quantum Quivers and Hall/Hole Halos},''
  JHEP {\bf 0210} (2002) 023
  [arXiv:hep-th/0206072].

\bibitem{Denef:2007vg}
  F.~Denef and G.W.~Moore,
  ``{Split States, Entropy enigmas, Holes and Halos},''
  [arXiv:hep-th/0702146].

\bibitem{KS}
  M. Kontsevich and Y. Soibelman,
``{Stability Structures, Motivic Donaldson-Thomas Invariants
and Cluster Transformations},'' [arXiv:0811.2435]


\bibitem{Gaiotto:2008cd}
  D.~Gaiotto, G.W.~Moore and A.~Neitzke,
``{Four-dimensional Wall-crossing via Three-dimensional
Field Theory},''
  Commun. Math. Phys. {\bf 299} (2010) 163
  [arXiv:0807.4723 [hep-th]].

\bibitem{Gaiotto:2009hg}
  D.~Gaiotto, G.~W.~Moore, A.~Neitzke,
  ``Wall-crossing, Hitchin Systems, and the WKB Approximation,''
[arXiv:0907.3987 [hep-th]].

\bibitem{Chen:2010yr}
  H.~-Y.~Chen, N.~Dorey, K.~Petunin,
  ``Wall Crossing and Instantons in Compactified Gauge Theory,''
  JHEP {\bf 1006 } (2010)  024.
  [arXiv:1004.0703 [hep-th]].

\bibitem{Kim:2011sc}
  H.~Kim, J.~Park, Z.~Wang and P.~Yi,
  ``Ab Initio Wall-Crossing,''
  JHEP {\bf 1109} (2011) 079
  [arXiv:1107.0723 [hep-th]].



\bibitem{Weinberg:2006rq}
  E.~J.~Weinberg and P.~Yi,
  ``Magnetic Monopole Dynamics, Supersymmetry, and Duality,''
  Phys.\ Rept.\  {\bf 438}, 65 (2007)
  [arXiv:hep-th/0609055].

\bibitem{Lee:2011ph}
  S.~Lee and P.~Yi,
  ``Framed BPS States, Moduli Dynamics, and Wall-Crossing,''
  JHEP {\bf 1104} (2011) 098
  [arXiv:1102.1729 [hep-th]].


\bibitem{Ritz:2008jf}
  A.~Ritz and A.~Vainshtein,
  ``Dyon dynamics near marginal stability and non-BPS states,''
  Phys.\ Lett.\  B {\bf 668}, 148 (2008)
  [arXiv:0807.2419 [hep-th]].

\bibitem{Sen:2011aa}
  A.~Sen,
  ``Equivalence of Three Wall Crossing Formulae,''
  arXiv:1112.2515 [hep-th].





\bibitem{Manschot:2010qz}
  J.~Manschot, B.~Pioline and A.~Sen,
  ``Wall Crossing from Boltzmann Black Hole Halos,''
  [arXiv:1011.1258 [hep-th]].
\bibitem{Chalmers:1996ya}
  G.~Chalmers, M.~Rocek and R.~von Unge,
  ``Monopoles in quantum corrected N=2 superYang-Mills theory,''
  arXiv:hep-th/9612195.

\bibitem{Mikhailov:1998bx}
  A.~Mikhailov, N.~Nekrasov and S.~Sethi,
  ``Geometric realizations of BPS states in N = 2 theories,''
  Nucl.\ Phys.\  B {\bf 531} (1998) 345
  [arXiv:hep-th/9803142].

\bibitem{Ritz:2000xa}
  A.~Ritz, M.A.~Shifman, A.I.~Vainshtein and M.B.~Voloshin,
``{Marginal Stability and the Metamorphosis of BPS
States},''
  Phys. Rev. D {\bf 63} (2001) 065018
  [arXiv:hep-th/0006028].

  \bibitem{Argyres:2001pv}
  P.C.~Argyres and K.~Narayan,
 ``{String Webs from Field Theory},''
  JHEP {\bf 0103} (2001) 047
  [arXiv:hep-th/0101114].

\bibitem{Wu:1976ge}
  T.T.~Wu and C.N.~Yang,
  ``{Dirac Monopole without Strings: Monopole Harmonics},''
  Nucl. Phys. B {\bf 107} (1976) 365.


\bibitem{Ivanov:1990jn}
  E.~A.~Ivanov and A.~V.~Smilga,
  ``Supersymmetric gauge quantum mechanics: Superfield description,''
  Phys.\ Lett.\  B {\bf 257} (1991) 79.

\bibitem{Berezovoj:1991ka}
  V.~P.~Berezovoj and A.~I.~Pashnev,
  ``Three-dimensional N=4 extended supersymmetrical quantum mechanics,''
  Class.\ Quant.\ Grav.\  {\bf 8} (1991) 2141.

  \bibitem{Maloney:1999dv}
  A.~Maloney, M.~Spradlin and A.~Strominger,
``{Superconformal Multi-Black Hole Moduli Spaces in Four
Dimensions},''
  JHEP {\bf 0204} (2002) 003
  [arXiv:hep-th/9911001].




\bibitem{Yi:1997eg}
  P.~Yi,
  ``Witten index and threshold bound states of D-branes,''
  Nucl.\ Phys.\  B {\bf 505}, 307 (1997)
  [arXiv:hep-th/9704098].

\bibitem{Green:1997tn}
  M.~B.~Green and M.~Gutperle,
  ``D Particle bound states and the D instanton measure,''
  JHEP {\bf 9801}, 005 (1998)
  [arXiv:hep-th/9711107].

\bibitem{Witten:1995im}
  E.~Witten,
  ``Bound states of strings and p-branes,''
  Nucl.\ Phys.\ B {\bf 460} (1996) 335
  [hep-th/9510135].


\bibitem{Gaiotto:2010be}
  D.~Gaiotto, G.W.~Moore and A.~Neitzke,
  ``{Framed BPS States},''
  [arXiv:1006.0146 [hep-th]].


\bibitem{Manschot:2011xc}
  J.~Manschot, B.~Pioline and A.~Sen,
  ``A Fixed point formula for the index of multi-centered N=2 black holes,''
  JHEP {\bf 1105}, 057 (2011)
  [arXiv:1103.1887 [hep-th]].

\bibitem{Lee:2012sc}
  S.~-J.~Lee, Z.~-L.~Wang and P.~Yi,
  ``Quiver Invariants from Intrinsic Higgs States,''
  arXiv:1205.6511 [hep-th].


\bibitem{Bena:2012hf}
  I.~Bena, M.~Berkooz, J.~de Boer, S.~El-Showk and D.~Van den Bleeken,
  ``Scaling BPS Solutions and pure-Higgs States,''
  arXiv:1205.5023 [hep-th].

\bibitem{Lee:2012na}
  S.~-J.~Lee, Z.~-L.~Wang and P.~Yi,
  ``BPS States, Refined Indices, and Quiver Invariants,''
  arXiv:1207.0821 [hep-th].

\bibitem{Manschot:2012rx}
  J.~Manschot, B.~Pioline and A.~Sen,
  ``From Black Holes to Quivers,''
  arXiv:1207.2230 [hep-th].

\end{thebibliography}
\end{document}